\documentclass[preprintnumbers,amsmath,amssymb]{revtex4}
\usepackage{graphicx}% Include figure files
\usepackage{dcolumn}% Align table columns on decimal point
\usepackage{bm}% bold math

\begin{document}

%\preprint{APS/123-QED}

\title{Demonstration of Temporal Distinguishability of Three and Four Photons\\with Asymmetric Beam Splitter}
% Force line breaks with \\

\author{Z. Y. Ou}
\affiliation{Department of Physics, Indiana University-Purdue
University Indianapolis, 402 N. Blackford
Street, Indianapolis, IN 46202, USA\\
Tel: 317-274-2125, Fax: 317-274-2393, Email: zou@iupui.edu}

 \author{B. H. Liu, F. W. Sun, Y. X. Gong, Y. F. Huang, and G. C. Guo}
 \affiliation{Key Laboratory of Quantum
Information, University of Science and Technology of China, CAS,
Hefei, 230026, the People's Republic of
 China}


\begin{abstract}
By using an asymmetric beam splitter, we observe the generalized
Hong-Ou-Mandel effects for three and four photons, respectively.
Furthermore, we can use this generalized Hong-Ou-Mandel
interferometer to characterize temporal distinguishability.

\end{abstract}

\maketitle

Indistinguishability is essential in achieving quantum
entanglement. As more particles get involved in quantum
information sciences \cite{bou}, their mutual indistinguishability
will play an paramount role in some of the quantum information
protocols. Thus assessing multi-particle indistinguishability has
become an increasingly urgent task.

Mathematically, the temporal indistinguishability of an $N$-photon
state is characterized \cite{ou} by the permutation symmetry of
its spectral wave function: $\Phi(\omega_1,\omega_2,...,\omega_N)
= \Phi(\mathbb{P}\{\omega_1,\omega_2,...,\omega_N\})$ with
$\mathbb{P}$ as the arbitrary permutation operator.  For the
two-photon case, the visibility in a Hong-Ou-Mandel interferometer
\cite{hom} characterizes the degree of permutation symmetry and
thus the degree of temporal indistinguishability \cite{wam,ser}.
Generalization of this scheme to an arbitrary $N$-photon case is
not trivial \cite{ou} and involves complicated scheme of
projection measurement \cite{sun,res,xiang}. In this paper, we
will demonstrate a simple scheme using only one adjustable
asymmetric beam splitter and multi-photon coincidence measurement
to characterize the temporal distinguishability of a multi-photon
state.

The idea is that for an input state of $|m, n\rangle$ to an
asymmetric beam splitter (Fig.1a) of transmissivity $T$ and
reflectivity $R$ ($T\ne R$), the probability $P_{out}(m, n)$ for
$m$ photons coming out in one port and $n$ photons in the other
port is zero with the appropriate values of $T$ and $R$, due to
multi-photon interference. Since the effect of interference
depends on photon distinguishability, we can use the visibility of
the above mentioned interference effect to characterize the
temporal distinguishability of the input state of $|m, n\rangle$
(Fig.1b,c).

It can be shown \cite{ou2} that for the case of $n=1$, $m=N$, we
have $T=NR=N/(N+1)$ for $P_{N+1}(N, 1) =0$ in the ideal case. But
for the non-ideal case, the visibility is simply $p/N$, where $p$
is the number of photons among the $N$ photons that are
indistinguishable from the single photon from the other port. This
result is exactly the same as the complicated projection
measurement scheme discussed in Ref.\cite{ou}. So as we scan the
delay between the single photon and the $N$ photons, we will see a
number of dips with visibility of $p/N$, as shown in Fig.1c. The
picture in Fig.1c will fully characterize the temporal
distinguishability of the $N$ photons in Fig.1b.

\begin{figure}[htb]
\begin{center}
\includegraphics[width= 7in]{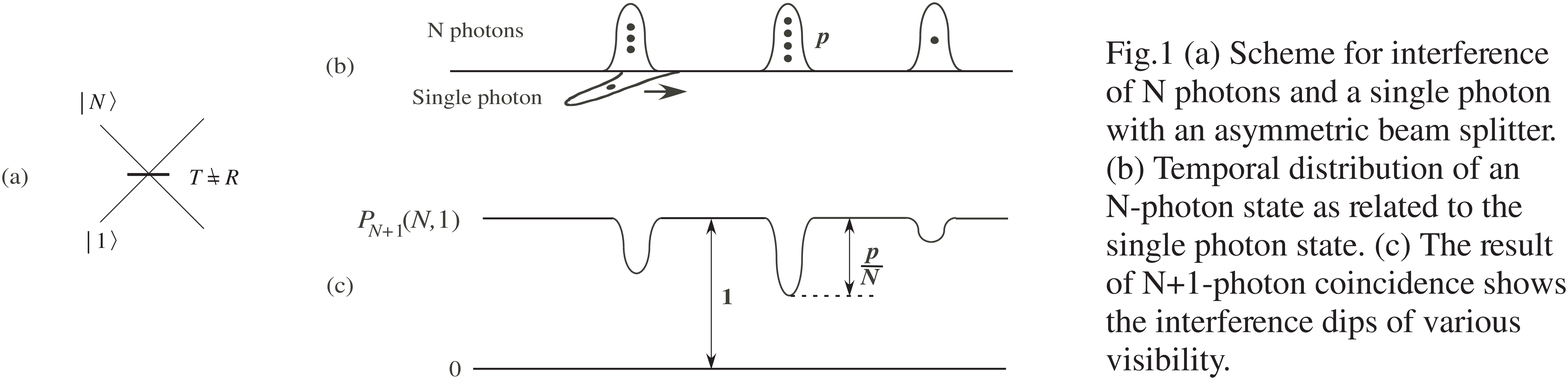}
\end{center}
\end{figure}

The case of $m=n=2$ can realized with $T=(3\pm\sqrt{3})/6,
R=(3\mp\sqrt{3})/6$. The visibility of different scenarios in this
case is exactly the same as the projection scheme in
Ref.\cite{ou}. But the case of other values of $m, n$ are
complicated.

Experimentally, we employ the state $|1_V, 2_H\rangle$ and
$|2_V,2_H\rangle$ in an asymmetry beam splitter made of half wave
plate (HWP) and polarization beam splitters (PBS) (Fig.2). With
proper rotation angle of the HWP, we may achieve $T=2/3$ for
three-photon case or $T=(3\pm\sqrt{3})/6$ for the four-photon
case. The two different detection schemes for three and four
photons are shown in the insets of Fig.2, respectively. The state
$|2_V,2_H\rangle$ can be easily generated from a type-II
parametric down-conversion process. But the state
$|1_V,2_H\rangle$ is not straightforward. We use a same scheme as
in Ref.\cite{liu} to produce a three-photon state with adjustable
temporal distinguishability.

We measure the three-photon coincidence (inset (a) of Fig.2) as a
function of the delay $\Delta$ between the two H-photons and the
single V-photon. There are two scenarios in this case: (a) the two
H-photons are indistinguishable and in the state of $2_H\rangle$
or (b) the two H-photons are well separated and become
distinguishable in the state of $1_H, 1'_H\rangle$. The results
for these to scenarios are shown in Fig.3a and Fig.3b,
respectively. Similarly, the four-photon result is shown in
Fig.3(c).

Both Fig.3(a) and Fig.3(c) show the typical Hong-Ou-Mandel dips
for three- and four-photon cases, respectively, with a large
visibility (85\% and 92\%). The double dip feature in Fig.3(b) is
due to temporal separation between the two H-photons with a proper
adjustment of their relative delay. The close to 50\% visibility
reflects the fact that the two H-photons are distinguishable. This
is consistent with the theory mention earlier. Thus this scheme
with an asymmetric beam splitter can characterize the degree of
temporal distinguishability of a multi-photon state by the
visibility of interference dips.

It should be pointed out that for $n=1, m=2$, the scenario of
indistinguishable two H-photons in Fig.3(a) was first discussed by
Wang and Kobayashi \cite{wan} and later realized experimentally by
Sanaka {\it et al.} \cite{san}.

\begin{figure}[htb]
\begin{center}
\includegraphics[width= 6in]{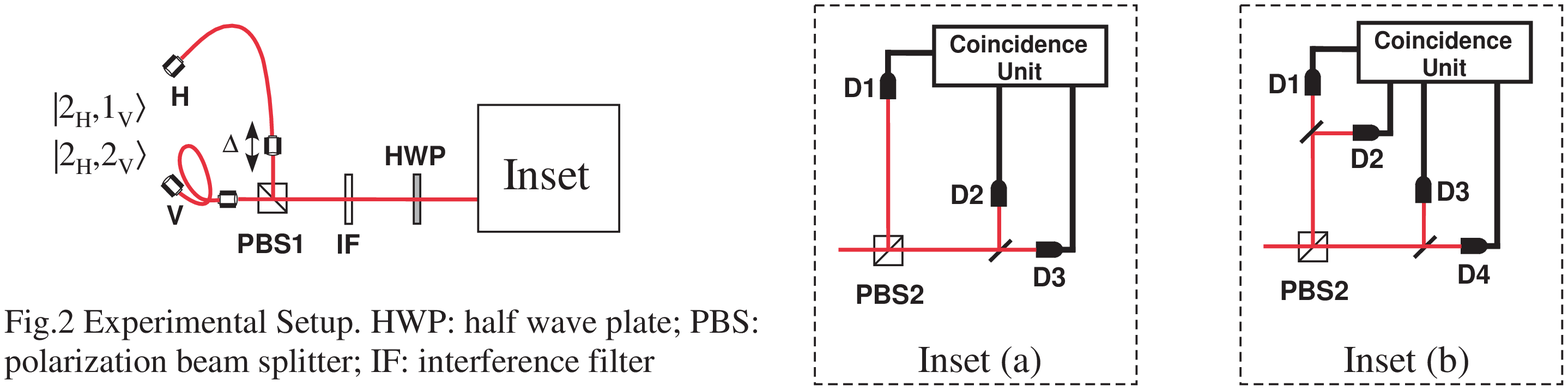}
\end{center}
\end{figure}

\begin{figure}[htb]
\begin{center}
\includegraphics[width= 7in]{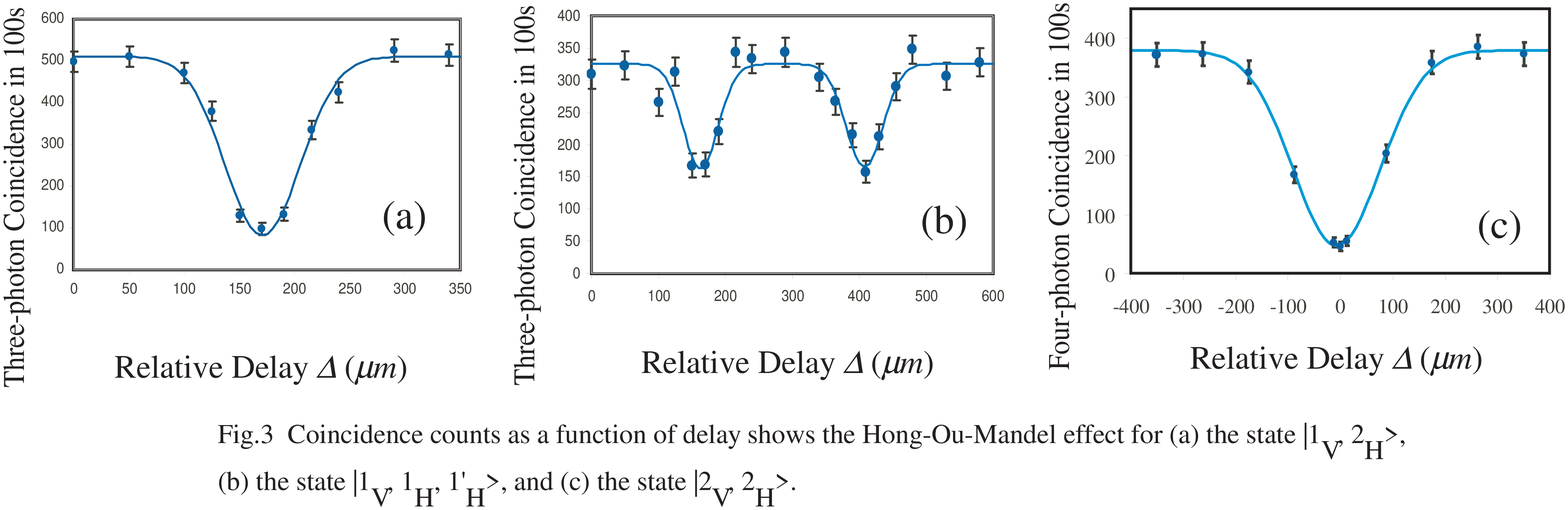}
\end{center}
\end{figure}

\end{document}